\documentclass[useAMS,usenatbib]{rmaa}

\begin{document}

\title{Detailed Atmospheric abundance analysis of the optical counterpart of the IR
source IRAS 16559-2957}

\author{R. E. Molina,\altaffilmark{1}
	and A. Arellano Ferro\altaffilmark{2}
\medskip }

\altaffiltext{1}{Laboratorio de Investigaci\'on en F\1sica Aplicada y Computacional,
Universidad Nacional Experimental del T\'achira, Venezuela.}
\altaffiltext{2}{Instituto de
Astronom\'{\i}a, Universidad Nacional Aut\'onoma de M\'exico,
M\'exico.}

\fulladdresses{
\item R. E. Molina: Laboratorio de Investigaci\'on en F\1sica Aplicada y
Computacional, Universidad Nacional Experimental del T\'achira,  Venezuela,
(rmolina@unet.edu.ve).
\item A. Arellano Ferro: Instituto de Astronom\1a, UNAM, Apartado Postal 70-264, 
04510, M\'exico, D. F., M\'exico (armando@astro.unam.mx).}

\shortauthor{}
\shorttitle{On the optical counterpart of IRAS~16559-2957}

\resumen{Hemos emprendido un an\'alisis detallado de las abundancias qu\1micas de
la contraparte \'optica de la fuente infrarroja IRAS~16559-2957 con el fin de
confirmar su posible naturaleza de estrella post-AGB. El an\'alisis de un gran
n\'umero de 
elementos, incluidos CNO y $^{12}$C/$^{13}$C, muestra que este objeto ha 
experimentado el primer dragado y parece estar a\'un en fase de RGB.}

\abstract{We have undertaken a detailed abundance analysis of the optical counterpart
of the IR source IRAS16559-2957 with the aim of confirming its possible post-AGB
nature.
The star shows solar metallicity and our investigation of a large number of elements
including CNO and $^{12}$C/$^{13}$C suggests that this object has experienced the
first dredge-up and it is likely still at RGB stage.}


\maketitle

\section{INTRODUCTION}

Post-AGB stars (PAGB's) are at a very advanced stage of evolution. They show 
peculiarities in the observed elements in their atmospheres (e.g. CNO and s-process
elements abundances) which are connected with the dredge-up episodes that 
occur as the star evolves up on the AGB.  PAGB's also have high mass loss 
($\geq$ 10$^{-6}$ M$_{\odot}$/yr) which obscures the central star and produces
emission in the IR and OH masers (Zijlstra et al. 1989). Consequently,
the IRAS color-color diagram ([25]-[60],[12]-[25]) and its IRAS variability has been 
widely used in the identification of PAGB and proto Planetary Nebulae (PPN)
candidates (van der Veen \& Habing 1988; Garc\1a-Lario et al. 1997; Su\'arez et al.
2006; Szczerba et al. 2012).

The identification of optical counterparts of IR and OH masers and IRAS sources is of
fundamental importance since it is in the optical spectrum that the abundance
anomalies of structural and evolutionary relevance can be detected and studied, such
as CNO abundances and those of s-process elements (e.g. La, Ba, Nd, Sr, Y, Zr). 

The IRAS 16559-2957 source has been included in several studies aimed to identify and
evaluate the evolutionary status of large lists of PAGB and PPN candidates. In the
finding chart provided by Hu et al. (1993) (their object 34) there is only one bright
star within the IRAS beam. According to Hu et al. it is a F5I star of $V \sim 13$
with H$\alpha$ filled in. The presence of a circumstellar envelope around a central
star was confirmed by the detection of strong OH maser at 1612 and 1665 MHz (Hu et
al. 1993). The radial velocity of OH maser profile varies between 57--70 km s$^{-1}$
(te Lintel Hekkert et al. 1991).

The source was included in the sample of PAGB and PN candidates studied by Su\'arez
et al. (2006) who find no optical counterpart. However, Ramos-Larios et al. (2009),
in their search of heavily obscured PAGB and PN stars found a near-IR counterpart in
the 2MASS Point Source Catalogue and an optical couterpart in the Digital Sky Survey,
which corresponds to a $V \sim 13.2$ star (see their finding chart on Fig. 3), in
agreement with Hu et al. (1993). Then the star has been "bona fide" catalogued as a
PAGB (e.g. Szczerba et al. 2007; SIMBAD \footnote
{http://simbad.u-strasbg.fr/simbad/sim-fid}data base).

With the aim of confirming the PAGB nature of the optical counterpart we have studied
a high resolution optical spectrum of the $V \sim 13.2$ star at RA~$\alpha =
16^{\mbox{\scriptsize h}} 59^{\mbox{\scriptsize m}} 08.2^{\mbox{\scriptsize s}}$,
Dec.~$\delta = -30^{\circ} 01\arcmin 40.3\arcsec$. In this paper, we present a
photospheric detailed chemical abundances analysis of this optical source and we
shall refer to the object as IRAS~16559-2957.

The paper is organized as follows: in  $\S$ \ref{sec:observ} the observations
and reductions are described, in $\S$ \ref{sec:param} we discuss the adopted atmospheric 
parameters and the reddening, in $\S$ \ref{sec:analy} we employed the equivalent
widths and
spectral synthesis approaches to derived the photospheric chemical abundances,
in $\S$ \ref{sec:result} we discuss our results, and in $\S$ \ref{sec:conclusion}
we summarize our conclusions.

\section{Observations and Reductions}
\label{sec:observ}

The spectrum of IRAS~16559-2957 ($V$=13.2) was obtained on February 27, 2008 with the
2D Coud\'e 
echelle spectrograph (Tull et al. 1995) on the 2.7m telescope at the McDonald Observatory 
giving $\sim$ 40,000 resolution and a wavelength coverage from 3900 to 10000 \AA~.
The spectrum was reduced using the IRAF astronomical routines. The equivalent widths (EW's)
were measured using the SPLOT task and their accuracy is generally
better than 10\% for spectra with S/N ratio larger than 50. We generally
restricted ourselves to unblended weak features and avoided using lines 
stronger than 200 m\AA.

We measure the heliocentric radial velocity using the Doppler shift of   
40 clean unblended absorption lines that cover a wide spectral range.
We derive a heliocentric radial velocity of  $-2.3 \pm 0.5$ km s$^{-1}$ (V$_{\rm LSR}$
= +7.1 km s$^{-1}$).

\section{Atmospheric parameters}
\label{sec:param}

Before we derive the elemental abundances it is necessary to estimate
the fundamental atmospheric parameters; effective
temperature, surface gravity and microturbulence velocity. We have approached 
the task as discussed in the following subsections.

\begin{table}
 \centering
 \begin{minipage}{85mm}
  \caption{Atmospheric parameters of IRAS~16559-2957.}
 \label{tab:table1}
\begin{tabular}{ccc}
  \hline
  \hline
\multicolumn{1}{l}{$T_{\rm eff,}$$_{sp}$}&
\multicolumn{1}{c}{log~$g$$_{sp}$}&
\multicolumn{1}{c}{$\xi_{t,}$$_{sp}$}\\
\multicolumn{1}{l}{(K)}&
\multicolumn{1}{c}{}&
\multicolumn{1}{c}{(km s$^{-1}$)}\\
\hline
     &      &       \\
4250 & 1.5  & 1.43  \\
\hline
\end{tabular}
\end{minipage}
\scriptsize{\begin{flushleft}
\end{flushleft}}
\end{table}

\subsection{Spectroscopy}
\label{sec:spectro}

An independent estimation of $T_{\rm eff}$, log~$g$ and the microturbulence velocity 
$\xi_{t}$ can be obtained from the behavior of the lines
of a well represented element (e.g. Fe),
and the {\tt MOOG} code (2009 version) by C. Sneden (1973), which is 
based on local thermodynamical equilibrium (LTE) calculations. We employed the 
atmospheric model collection of Castelli \& Kurucz (2003).

We have therefore followed the standard procedure for temperature and gravity 
determinations. For temperature, it is required that the Fe~I abundances are
independent of the low excitation potential of the lines. The gravity can be derived
by requiring that the Fe~I and Fe~II lines yield the same abundance.  This
procedure led to values $T_{\rm eff}= 4250\pm200$K and log~$g$=1.5$\pm0.25$. 
This effective temperature does not correspond to a star of spectral type F5I  but
it is more typical of an early K-type star.

\begin{figure}
\begin{center}
  \includegraphics[width=\columnwidth]{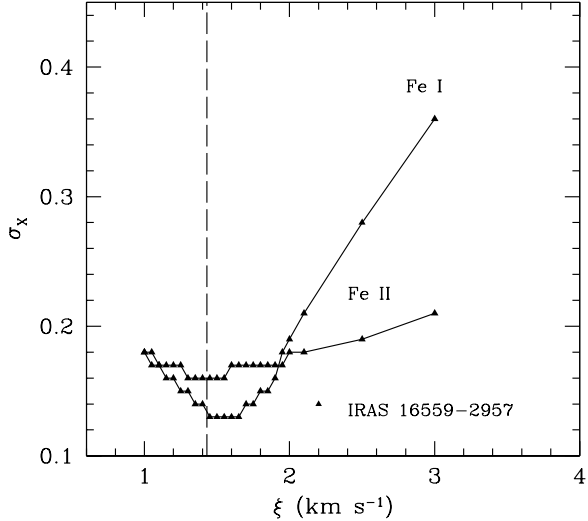}
  \caption{The standard deviation of Fe I and Fe II 
abundances as a function of microturbulence velocity for IRAS~16559-2957. 
The dotted vertical line corresponds to the value of the adopted microturbulence velocity.}
  \label{fig:figure1}
\end{center}
\end{figure}

The microturbence is estimated by requiring that the derived abundances 
are independent of the line strengths for a given specie. Generally, lines of Fe~II
are preferred.
It  is known  that Fe~II lines are not seriously affected
by departure from LTE (Schiller \& Przybilla 2008). On the contrary Fe~I lines
are affected by NLTE effects (Boyarchuck et al. 1985; Th\'evenin \& Idiart 1999).
For cool stars however not many Fe~II lines are available.
The microturbulence velocity, is therefore determined using the error diagram for 
Fe~I or Fe~II (depending on the number of measured lines). 
For a given model, we compute the dispersion in the Fe abundances 
over a range in the microturbulence velocity using the method of
Sahin \& Lambert (2009). Fig.~\ref{fig:figure1} shows the standard deviation
$\sigma_{x}$ as a function of microturbulence velocity $\xi_{t}$
for Fe I and Fe II lines. From the minimum error we estimate $\xi_{t}$ =1.43 km
s$^{-1}$.

\subsection{Determination of E(B-V)}
\label{sec:photom}
The initial effective temperature of IRAS~16559-2957 could also be estimated
from $JHK$ photometry  available in the work of Cutri et al. (2003) and the $T_{\rm
eff}$ calibrations of Alonso et al. (1999) from the $J-H$ and $V-K$ colours. However,
a previous knowledge of the reddening would be necessary. Instead we have used the
calibrations to estimate the total extinction.
The insterstellar extinction maps of Schlegel et al. (1998) lead to a value
$E(B-V)=0.38$, and after applying the correction recommended by Bonifacio et al.
(2000) for exceses larger than 0.15, we find $E(B-V)=0.28$.
Since the star is affected by circumstellar material responsible of
the OH masser and IR excess, we have increased the value of $E(B-V)$ until the
calibrations of Alonso et al. (1999) produce a value of $T_{\rm eff}$ $\sim$4250K,
estimated
spectroscopically. The total, interstellar plus circumstellar, reddening is
$E(B-V)=0.81$, leaving $E(B-V)=0.53$ as our estimation of the circumstellar
contribution.

The basic photometric and spectroscopic atmospheric parameters 
are resumed in Table~\ref{tab:table1}. The adopted values for the atmospheric
parameters are indicated by the lower subscript "sp". 

The models of Castelli \& Kurucz (2003) are calculated in steps of  $\pm$ 250~K, 
$\pm$0.5 and $\pm$0.25 km s$^{-1}$ in 
 $T_{\rm eff}$, log~$g$ and $\xi_{t}$ respectively. We have used however a set of
interpolated models in steps of $\pm$100~K, $\pm$0.25 and $\pm$0.20 km s$^{-1}$.
The sensitivity of the derived abundances to the uncertainties of
atmospheric parameters are given in Table \ref{tab:table3} were
we present changes in abundances caused by varying atmospheric
parameter by the above amounts with respect to the chosen model.

\section{Atmospheric Abundances}
\label{sec:analy}

\subsection{Equivalent Width Analysis}
\label{sec:abund}

To calculate the atmospheric abundances we have used the 2009 version of the
code {\tt MOOG} (Sneden 1973) which performs LTE line analysis. We use the grid
ATLAS9 of plane-parallel model atmosphere computed by Castelli \& Kurucz (2003).
Chemical abundances of well represented elements were
determined through the analysis of equivalent widths using
the \emph{abfind} driver in MOOG code. 

The oscillator strength or ${\rm gf}$ value is an important atomic
parameter that affects the abundance calculations.
The ${\rm gf}$ values can be obtained from numerous sources,
and the uncertainties vary from element to element. For example, experimental values
for Fe~I and Fe~II of high accuracy, between 5 to 10\%, are
available for a large fraction of lines. For other Fe-peak
elements, errors in their ${\rm gf}$ values may range between 10 to 25\%. For neutron-capture elements the
 accuracy of recent estimates are in 10\% to 25\% range.
An extensive list of  ${\rm gf}$ values for all important elements can be
found in  Sumangala Rao, Giridhar \& Lambert (2012).

\begin{table}
\caption{Sensitivity of abundances relative to the uncertainties in the model
parameters for IRAS~16559-2957.}
\label{tab:table3}
\begin{center}
\begin{tabular}{lrrr}
\noalign{\smallskip}
\hline
\hline
\noalign{\smallskip}
\noalign{\smallskip}
\multicolumn{1}{l}{Species}&
\multicolumn{1}{c}{$\Delta$ $T_{\rm eff}$}&
\multicolumn{1}{c}{$\Delta$ $\log$~$g$}&
\multicolumn{1}{c}{$\Delta$ $\xi_{t}$}\\
\multicolumn{1}{l}{}&
\multicolumn{1}{c}{$+$100~K}&
\multicolumn{1}{c}{$+$0.25}&
\multicolumn{1}{c}{$+$0.2 km s$^{-1}$}\\
           \noalign{\smallskip}
            \hline
            \noalign{\smallskip}
C   & $+0.02$ & $-0.05$ & $0.00$  \\
N    & $-0.05$ & $-0.08$ & $0.00$  \\
O  I & $-0.02$ & $-0.11$ & $+0.01$ \\
Li I & $-0.12$ & $+0.05$ & $0.00$  \\
Na I & $-0.10$ & $0.00$ & $+0.10$ \\
Mg I & $-0.01$ & $-0.04$ & $+0.09$ \\
Al I & $-0.07$ & $-0.03$ & $+0.07$ \\
Si I & $+0.07$ & $-0.12$ & $+0.05$ \\
K  I & $+0.10$ & $+0.10$ & $+0.10$ \\
Ca I & $-0.10$ & $+0.01$ & $+0.15$ \\
Sc II& $0.00$  & $-0.15$ & $+0.10$ \\
Ti I & $-0.17$ & $-0.01$ & $+0.09$ \\
V II &  $-0.15$ & $-0.03$ & $+0.17$ \\
Cr I &  $-0.12$ & $+0.03$ & $+0.13$ \\
Cr II& $+0.09$ & $-0.18$ & $+0.05$ \\
Mn  I& $+0.05$ & $+0.05$ & $0.00$  \\
Fe  I& $-0.01$ & $-0.08$ & $+0.13$ \\
Fe II& $+0.14$ & $-0.24$ & $+0.07$ \\
Co  I& $-0.01$ & $-0.10$ & $+0.11$ \\
Ni I & $0.00$  & $-0.10$ & $+0.16$ \\
Cu I & $+0.10$ & $0.00$  & $+0.20$ \\
Zn I & $+0.12$ & $-0.16$ & $+0.05$ \\
Rb I & $-0.15$ & $0.00$  & $0.00$  \\
Y  I & $-0.19$ & $0.00$  & $+0.07$ \\
Y  II& $-0.01$ & $-0.13$ & $+0.07$ \\
Zr I & $-0.10$ & $+0.05$ & $0.00$  \\
Ba II & $0.00$  & $-0.10$ & $+0.25$ \\
La II& $-0.02$ & $-0.12$ & $+0.08$ \\
Nd II& $-0.03$ & $-0.11$ & $+0.02$ \\
Eu II& $0.00$  & $-0.12$ & $0.00$  \\
            \noalign{\smallskip}
            \hline
            \noalign{\smallskip}
\end{tabular}
\end{center}
\end{table}

The sensitivity of the derived abundances to the uncertainties in the
model atmosphere parameters $T_{\rm eff}$, log~$g$ and
$\xi_{t}$ are summarized in Table~\ref{tab:table3}. We present abundance variations 
caused by changes in atmospheric parameters of 100~K, 0.25 dex 
and 0.2 km s$^{-1}$ respectively, relative to the selected model. We have 
evaluated the total error in the abundance of a given element,
by taking the square root of the summation of the
squares of the errors associated to $T_{\rm eff}$, log~$g$ and
$\xi_{t}$. From the uncertainties listed in Table~\ref{tab:table3},
we find the total absolute uncertainty to be ranging from 0.05 for C to
0.29 for Fe II. 

\subsection{Spectrum Synthesis Analysis}
\label{sec:synth}

In this section we discuss the determination of abundances of light
elements CNO. Some lines that are blended or affected by the
hyper-fine structure were analyzed by spectral synthesis using the \emph{synth}
driver in the MOOG code. We have validated line list used as follows.
We first have used the linelist of the corresponding spectral region and fitted the
synthesized spectrum with the solar flux atlas of Kurucz et al. (1984). Since many
molecular lines are not very strong in the solar spectrum, we further repeated the
procedure with the spectral atlas of Arcturus (Hinkle, et al. 2000) to corroborate the
input atomic and molecular data. There are still a few unidentified features that
could contribute to the uncertainties.

\subsubsection{Carbon}
\label{sec:carbon}

The carbon abundances were derived from the C$_{2}$ Swan band lines at
$\lambda$4730--4750 \AA~ and $\lambda$5160--5167 \AA.
An alternate region containing CH molecules located at 4300-4310 \AA~ (G-band) has also been used
for this purpose. 
The C$_{2}$ lines have been shown to be useful carbon abundance indicators in carbon stars
(Zamora et al. 2009) and in Galactic R Coronae Borealis stars (Hema et al. 2012).
 Since IRAS~16559-2957 is an O-rich star (C/O$\sim$0.11), C being locked up in
molecules is important, hence
allowance was made for C in molecules CH, C2, CN and CO at the
time of synthetizing C$_{2}$ and CH.
For IRAS~16559-2957 the C abundance has a mean value $\varepsilon$(C)=7.75$\pm$0.20,
i.e. the star is carbon deficient. 

\begin{figure}
\begin{center}
  \includegraphics[width=\columnwidth]{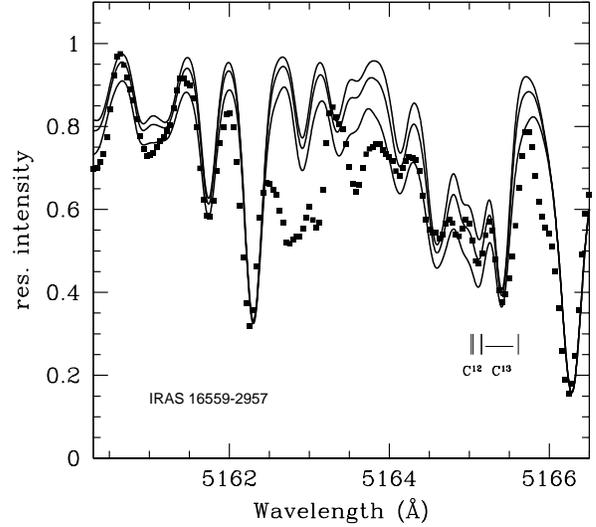}
  \caption{Observed (filled squares) and synthetic (solid lines) spectra in the
$\lambda$5165 \AA~C$_{2}$ band region for IRAS~16559-2957 star. The synthetic spectra
correspond, from top to bottom, to $\log$ $\varepsilon$(C)= 7.95, 7.75, and 7.55.}
\label{fig:figure2}
\end{center}
\end{figure}

Figure~\ref{fig:figure2} shows an example of the observed and synthetic spectra 
in the region around $\lambda$5165 \AA~.

\subsubsection{Nitrogen}
\label{sec:nitrog}

The nitrogen abundances were derived from the CN molecular lines also known as
the CN red system because the molecular bands occurs at $\lambda$ $\ge$ 5800 \AA.
In this study we use the CN(5,1) molecular line at $\lambda$6332.18 \AA.
The CN(5,1) feature is affected by blends with two lines
at $\lambda$6331.95 \AA, one due to Si I and the other due to
Fe II. Both features are taken into account in the synthesis.
In order to derive the abundance of N is necessary to consider the C abundance
previously obtained from Swan bands.

\begin{figure}
\begin{center}
  \includegraphics[width=\columnwidth]{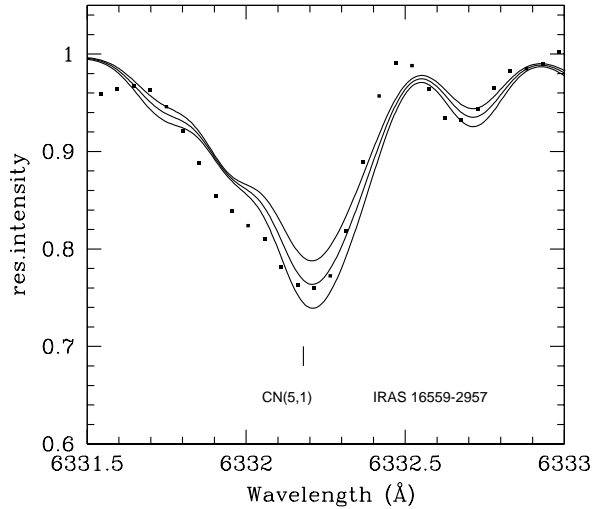}
  \caption{Observed (filled squares) and synthetic (solid lines) spectra
of the CN(5,1) band. The synthetic spectra correspond,
from top to bottom to $\log$ $\varepsilon$(N)= 9.10, 9.20, and 9.30.}
\label{fig:figure3}
\end{center}
\end{figure}

Figure~\ref{fig:figure3} shows the observed and synthetic spectra of
IRAS~16559-2957 in the region $\lambda$6331--6333 \AA. We estimate a value of $\log$
$\varepsilon$(N)=9.2$\pm$0.3 from $\lambda$6332 \AA~ region. This value is abnormally
high and is estimated with larger uncertainty probably due to the fact that the
feature is heavily blended and the region contains unidentified lines.       

\subsubsection{Oxygen}
\label{sec:oxyg}

Unfortunately, the [O I] forbidden line at 6300.304 \AA~ region is not present in the
observed spectrum of IRAS~16559-2957 as it falls in an inter order gap. The [O
I] line at 5577.3 gives a value $\varepsilon$(O)=9.0, however, it is blended with a
C$_2$ feature and we opted for not using it. Thus, we determined the oxygen abundance
from the equivalent width of the line 6363.78 \AA.
However, the derived value of $\varepsilon$(O)=8.7  should be considered an upper
limit since this line seems
contaminated with either nearby metallic lines or molecular CN.

\subsubsection{$^{12}$C/$^{13}$C}
\label{sec:isotop}

\begin{figure}
\begin{center}
  \includegraphics[width=\columnwidth]{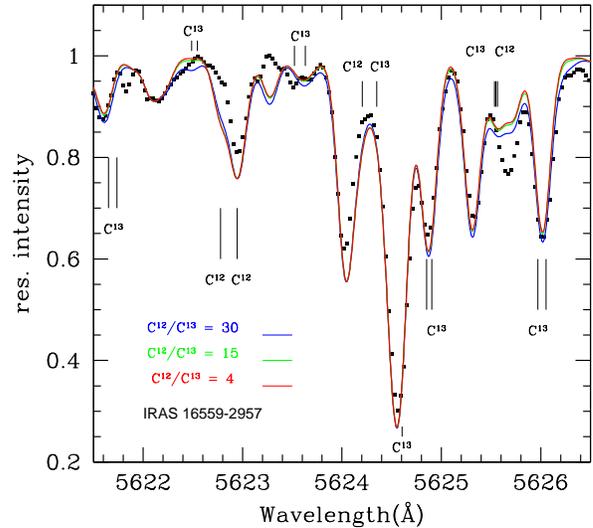}
  \caption{Determination of $^{12}$C/$^{13}$C isotope ratio for IRAS~16559-2957. 
Observed (filled squares) and synthetic (solid lines) spectra in
$\lambda$5622--28 \AA~. Our best estimate is  $^{12}$C/$^{13}$C = 15 (green).}
\label{fig:figure4}
\end{center}
\end{figure}

The $^{12}$C/$^{13}$C isotope ratio was determined using the features
of $^{12}$CN and $^{13}$CN molecules in the 5622-26 \AA~ spectral
region (Fig.~\ref{fig:figure4}). We derived a value
of the $^{12}$C/$^{13}$C isotope ratio equal to 15.

\subsubsection{Hyper-fine structure considerations}
\label{sec:other}

Hyper-fine structure (hfs) has the effect of de-saturating strong lines; hence, 
it is important to consider hyper-fine components when performing abundance 
analysis of strong lines (McWilliam et al. 1995). For weak unsaturated lines the hfs
treatment is not
necessary.

In this analysis we have adopted the hfs data of Spite et al. (1989) and
Prochaska \& McWilliam (2000) for Sc II lines, Prochaska \& McWilliam (2000)
and Allen \& Porto de Mello (2011) for Mn I lines, Mucciarelli et al. (2008)
for Eu II line, McWilliam (1998) for Ba II lines,
Allen \& Porto de Mello (2011) for Cu I lines, Reddy et al. (2002) for Li I doublet
at $\lambda$6707.7 \AA~ and Lambert \& Luck (1976) for Rb I at 7800.3 \AA.

\begin{table*}
\centering
\setlength{\tabnotewidth}{0.4\columnwidth}
\setlength{\tabcolsep}{0.70\tabcolsep} \tablecols{9}
 \caption{CNO abundances for IRAS~16559-2957.}
 \label{tab:table4}
\begin{tabular}{ccccccc}
  \hline\hline
\multicolumn{1}{c}{C/O}&
\multicolumn{1}{c}{C/N}&
\multicolumn{1}{c}{$^{12}$C/$^{13}$C}&
\multicolumn{1}{c}{[C/Fe]}&
\multicolumn{1}{c}{[N/Fe]}&
\multicolumn{1}{c}{[O/Fe]}&
\multicolumn{1}{c}{[Fe/H]}\\
 \hline
0.11$\pm$0.08 & 0.04$\pm$0.03 & 15 & $-$0.68$\pm$0.27 & $+$1.38$\pm$0.42
& 0.00$\pm$0.19 & $+$0.04$\pm$0.19 \\
\hline
\end{tabular}
\end{table*}

\section{Results and Discussion}
\label{sec:result}

\subsection{Derived Elemental Abundances}

In Table~\ref{tab:table5} we present the calculated elemental abundances
of IRAS~16559-2957 obtained as discussed in the previous sections\footnote
{We adopt the usual notation [X/H] = $\log N(X)/N(H)_{*} - \log N(X)/N(H)_{\odot}$,
where$\log N(H) \equiv 12$ is the hydrogen abundance by number.}. The solar
abundances are taken from Asplund et al. (2005). 

\begin{table}
\caption{Elemental abundances of IRAS~16559-2957.
}
\label{tab:table5}
\begin{center}
\begin{tabular}{lcrcrr}
\noalign{\smallskip}
\hline \hline
\noalign{\smallskip}
\noalign{\smallskip}
\multicolumn{1}{l}{Species}&
\multicolumn{1}{c}{$\log \epsilon_{\odot}$}&
\multicolumn{1}{l}{[X/H]}&
\multicolumn{1}{l}{s.d.}&
\multicolumn{1}{c}{N}&
\multicolumn{1}{r}{[X/Fe]}\\
           \noalign{\smallskip}
            \hline
            \noalign{\smallskip}
C(C$_{2}$) & 8.39  & $-0.64$ &      & syn & $-0.68$ \\
N(CN) & 7.78 & $+1.42$ &           & syn & $+1.38$ \\
O  I & 8.66 & $+0.04$ &           & 1   & $0.00$ \\
Li I & 1.05 & $-0.07$ &           & syn & $-0.11$ \\
Na I & 6.17 & $+0.47$ & $\pm$0.19 & 3   & $+0.43$ \\
Mg I & 7.53 & $+0.24$ & $\pm$0.16 & 2   & $+0.20$ \\
Al I & 6.37 & $+0.10$ & $\pm$0.14 & 4   & $+0.06$ \\
Si I & 7.51 & $+0.05$ & $\pm$0.13 & 8   & $+0.01$ \\
K  I & 5.08 & $-0.28$ &           & syn & $-0.32$ \\
Ca I & 6.31 & $+0.01$ & $\pm$0.09 & 6   & $-0.03$ \\
Sc II& 3.05 & $-0.15$ &           & syn & $-0.19$ \\
Ti I & 4.90 & $-0.27$ & $\pm$0.12 & 11  & $-0.31$ \\
V  I & 4.00 & $+0.14$ & $\pm$0.15 & 14  & $+0.10$ \\
Cr I & 5.64 & $-0.13$ & $\pm$0.08 & 2   & $-0.17$ \\
Cr II& 5.64 & $-0.06$ & $\pm$0.13 & 3   & $-0.10$ \\
Mn  I& 5.39 & $-0.37$ &           & syn & $-0.41$ \\
Fe  I& 7.45 & $+0.06$ & $\pm$0.13 & 42  &  \\
Fe II& 7.45 & $+0.03$ & $\pm$0.16 & 9   &  \\
Co  I& 4.92 & $+0.35$ & $\pm$0.17 & 4   & $+0.31$ \\
Ni I & 6.23 & $+0.28$ & $\pm$0.12 & 13  & $+0.24$ \\
Cu I & 4.21 & $-0.71$ &           & syn & $-0.75$ \\
Zn I & 4.60 & $+0.08$ &           & 1   & $+0.04$ \\
Rb I & 2.60 & $-0.20$ &           & syn & $-0.23$ \\
Y  I & 2.21 & $-0.27$ &           & 1   & $-0.31$ \\
Y II & 2.21 & $-0.37$ &           & 1   & $-0.41$ \\
Zr I & 2.59 & $-0.19$ &           & syn & $-0.23$ \\
Ba II& 2.17 & $-0.62$ &           & syn & $-0.66$ \\
La II& 1.13 & $+0.34$ & $\pm$0.25 & 2   & $+0.30$ \\
Nd II& 1.45 &$-0.63$ &           & 1   & $-0.67$ \\
Eu II& 0.52 &$-0.22$ &           & syn & $-0.26$ \\
            \noalign{\smallskip}
            \hline
            \noalign{\smallskip}
\end{tabular}
\end{center}
\end{table}

\subsection{Carbon, Nitrogen and Oxygen}
\label{sec:cno}

In order to infer the evolutionary status of IRAS~16559-2957 the determination of CNO
abundances is required. Table~\ref{tab:table4} shows the C/O, C/N and
$^{12}$C/$^{13}$C ratios, the abundance of CNO relative to the iron abundance.
An average iron abundace of [Fe/H] = $+0.04\pm0.19$ was obtained from 42 Fe I
and 9 Fe II lines.

The peculiarity present in the CN abundances is the 
deficiency of C ([C/Fe]= $-0.68$) and the enhancement of N ([N/Fe]=+1.38). Our
value $\sum$CNO/Fe= +0.16$\pm$0.2 is nearly solar, as expected before the third
dredge up (Iben 1964) and being O abundance also solar, the C defficiency and
its uncertainty impact on the N abundance. The uncertainty in the C abundance $\pm0.2$
indicates that C might be underestimated by a factor of 1.6. This and the non-LTE
effect may produce an overestimation of N by as much as 0.3-0.4 dex, however the N
enhancement is still significant.

Strong C deficiency and N enhancement are the consquence of CN-cycled
material dredged up to the suface from the He-burning shell (first dredge up)
in stars upon ascending to the red giant branch. 
It has also been observed in core He-burning stars (Tautvai$\check{s}$ien\'e et al.
2001) and in RHB stars (Afsar et al. 2012). 

In massive post-AGB stars ($M > 4M_{\odot}$) N is also enhanced as in the hot bottom
of the convective envelope carbon produced by helium burning is converted into
nitrogen (Lattanzio \& Wood 2004; Giridhar 2011). This keeps the C/O ratio below unity
and the star does not become C-rich (Boothroyd et al. 1993). This Hot Bottom Burning
process also features efficient proton capture and hence s-process element
enhancements. 

IRAS~16559-2957 is an O-rich (C/O~$\sim$0.11) and is not s-process enriched.
The high nitrogen abundance [N/Fe] = +1.38 clearly exceeds the
mean observed values in giant ([N/Fe] = +0.35) and dwarf ([N/Fe] = +0.30) stars
(Luck 1991, Reddy et al. 2006). The O 
abundance, [O/Fe]=0.0, is similar to those observed in giant stars ([O/Fe] $\sim$
+0.08) (Luck 1991; Mishenina et al. 2006; Lambert \& Ries 1981; Bensby et al. 2003).

\subsection{Other heavy elements}
\label{sec:heavy}

Other relevant abundances are addressed in this section.
The Li abundance shows depletion. Theoretical predictions suggests that the stellar Li
abundance, if starting from the present interstellar medium abundance $\log$
$\varepsilon$(Li) = 3.3, would be down to around 1.5 after the first dredge up. Hence
G and K giants are expected to have low lithium abundances (Brown et al. 1989). The
value found for IRAS~16559-2957, $\log$ $\varepsilon$(Li) = 0.98$\pm$0.13 is therefore
in reasonable agreement with the first dredge-up calculations and  suggests
that surface Li abundance has been depleted already during the main sequence
evolution.

IRAS~16559-2957 shows relative enrichment of sodium ([Na/Fe]=+0.43). A possible source
of error as explained by Mishenina et al. (2006) is non-LTE effect. Our Na abundance
was calculated from three lines: 5682, 6154 and 6160 \AA\AA, with equivalent widths
$\sim$ 135 m\AA~ for which the corrections would be
in the +0.10 to +0.15 range, hence the relative enhancement of Na is real and not
caused by the neglect of non-LTE treatment.

The abundances of $\alpha$ elements are, within the uncertainties, of solar value as
they are  in dwarf and giant stars of the thin disk (Reddy et al. 2003, 2006; Luck \&
Heiter 2007).

The abundances of iron-peak elements (Sc, V, Mn, Fe, Co, Ni) in IRAS~16559-2957 
are consistent with the calculated trends by Takeda et al. (2008)
and Wang et al. (2011) for field G-giants, Reddy et al. (2003; 2006) for FG-dwarfs 
and Reddy et al. (2012) for red giant members of open cluster. 

The copper underabundance [Cu/Fe]=$-$0.75 calls our attention. This is comparable
to values in metal-poor field stars (Sneden et al. 1991). Expected values for [Cu/Fe]
in open cluster giants are between -0.11 to -0.23 while for normal giants of the thin
disk the value is expected to be +0.01$\pm$0.13 (Reddy et al. 2012, Luck \& Heiter
2007). This value defies a simple explanation. 

The abundance of rubidium was obtained from spectral synthesis using the Rb I line
at 7800 \AA. 
Our synthesis profile takes into account the hyper-fine structure of the $^{85}$Rb and
$^{87}$Rb 
isotopes (Lambert \& Luck 1976). The fit to this line includes the Si I line at
7799.99 \AA~ for which we took its ${\rm gf}$ value from Gustafsson et al. (2008) and
we checked the atomic data of the spectral region by fitting the Arcturus spectrum
(Hinkle et al. 2000).
The obtained abundance value [Rb/Fe] = $-$0.23 $\pm$ 0.19 can be compared with the
results 
of Carney et al. (2005) for four field giant stars in the direction of the southern warp of the 
Galactic disk ($<$[Rb/Fe]$>$ = $+$0.12 $\pm$ 0.03). We find that the difference is
hardly significant. Our value is also consistent with the mean abundance ratio of six
giant stars in the old open clusters Be 20 and Be 29
in the outer Galactic disk ($<$[Rb/Fe]$>$ = $-$0.09 $\pm$ 0.12) and of three giant
stars in the old open 
cluster M67 ($<$[Rb/Fe]$>$ = $-$0.27 $\pm$ 0.05) (Carney et al. 2005).

In the spectrum of IRAS~16559-2957 we have detected lines of the neutron capture
elements Y, Zr, Ba, La, Nd, and Eu. The derived abundances of these elements  are
similar to those observed in field G-giants (Takeda et al. 2008; Wang et al. 2011), in
FG-dwarfs (Reddy et al. 2003; 2006) and in red giant members of open cluster (Reddy et
al. 2012). However we note that Ba and Nd are significantly diminished relative to
those observed in dwarf and giant stars.

\subsection{Discussion}
\label{sec:discuss}

The chemical diversity in the central stars of PAGB's is very large. They
spread a large range of temperatures and masses below 8$M_{\odot}$. Different families
are described in detail in the thematic review by van Winckel (2003) and more
recently by Garc\'ia-Lario (2006) and Giridhar (2011). There are two known types of
O-rich PAGB's; the so called $classical$, optically bright which exhibit low
metallicities ([Fe/H] between -1 and -0.3) and display a double peak in their
Spectral Energy Distribution (SED), the optical one from the central star and the
near-IR one from the circumstellar material. They do not display s-process
enhancements and seem to be the successors of AGB stars that do not experience an
efficient third dredge-up (van Winckel 2003). 

The second group is formed by more massive stars ($M>4M_{\odot}$) where the Hot Bottom
Burning burns $^{12}$C into $^{13}$C and $^{14}$N and the products are
brought up to the surface via the third dredge-up; this can keep the C/O ratio below
unity, mantain the $^{12}$C/$^{13}$C ratio near the CN equilibrium value of 3.5 and
dramatically enhance $^{14}$N as explained by Lattanzio (2003). The stars in the later
group are surrounded by dusty molecular shells responsible of IR emission and
CO and OH masers.

The optical counterpart of IRAS~16559-2957 is an O-rich star, and its most prominent
spectral properties found in the present paper are: mild iron defficiency [Fe/H]=
$+0.04\pm0.19$, C/O=0.11, $^{12}$C/$^{13}$C=15, and no signs of s-process
enhancements. In fact an indicator of the influence of proton flux is the ratio of
heavy to light s-process elements or [hs/ls] (Luck \& Bond 1991). In IRAS~16559-2957
we calculate [hs/ls]=$-0.04$, where we have taken the abundances of Ba, La and Nd as
representatives of the heavy s-process elements and Y and Zr a representative of the
light ones. Typical value in s-processed enhanced PAGB's is [hs/ls]$\sim$10 (Reddy et
al. 2002). 

The isotope ratio $^{12}$C/$^{13}$C is not by itself an indicator of evolution from
AGB to PAGB stages and its value change dramatically between C-rich and O-rich stars.
In C-rich stars a lower limit of $^{12}$C/$^{13}$C $> 20$ has been found by Bakker et
al. (1997) using circumstellar molecular CN red lines, and $^{12}$C/$^{13}$C $> 25$
found
by Reddy et al. (2002) using $^{12}$CN and $^{13}$CN features in the near-IR. Values
in
the range $30<^{12}$C/$^{13}$C$<70$ were found in a sample of carbon stars in the
Galactic disk by Lambert et al. (1986).
In O-rich stars however, the first dredge up reduces the surface $^{12}$C/$^{13}$C
ratio and in low mass stars the inneficient third dredge up mantains the ratio low.
More massive stars with an intense Hot Bottom Burning (HBB) also mantain the isotope
ratio low.
$^{12}$C/$^{13}$C ratios as low as 3-5 has been found in O-rich planetary nebulae
(Rubin et al. 2004). According to Lattanzio \& Forestini (1999) low values of
$^{12}$C/$^{13}$C are strong evidence of HBB while high values are indication of
$^{12}$C dredge up without HBB and intermediate values are harder to interpret.

Thus, although not an indication of evolution in our O-rich star, our value of
$^{12}$C/$^{13}$C = 15 is consistent with first dredge up predictions, accompanied of
C decrese and N enhancement (Iben 1965). It is also comparable to the mean ratio
$^{12}$C/$^{13}$C = 16 in the sample of mildly metal poor giants of Cottrell \& Sneden
(1986). Giants in older open clusters, with lower turn-off
mass, exhibit values between 10 and 20 (Gilroy \& Brown 1991; Smiljani et
al. 2009). The clump giant stars in open clusters also shows low
values of $^{12}$C/$^{13}$C ratios suggesting a non-standard mixing episode occurring 
sometime along the upper RGB (Tautvai$\check{s}$ien\'e et al. 2010).

Lithium in IRAS~16559-2957 is depleted. Lithium can be produced in AGB
stars through the Cameron-Fowler Mechanism (Cameron \& Fowler 1971) and large
enhancements have been observed in AGB's in the Magellanic Clouds (Smith \& Lambert
1990). This, which is the consequence of HBB, seems to work only in intermedite mass
stars ($M>5M_{\odot}$) (Lattanzio \& Forestini 1999). 

In IRAS~16559-2957 it was noted that the Na is enriched showing that the surface
abundance of this element has suffered changes at the AGB, likely caused by proton
capture on $^{22}$Ne during the HBB phase as explained by Lattanzio (2003).

\section{Conclusions}
\label{sec:conclusion}

 We present the results of a detailed atmospheric abundance analysis of the star 
IRAS~16559-2957, previously classified as a post-AGB. The analysis is based on a high
resolution spectrum of the V$\sim$13 mag object at the coordinates of
IRAS~16559-2957 and includes 27 elements. The photospheric chemical 
abundances are derived by spectral synthesis and from the equivalent widths of
selected spectral lines.

Our spectrum is quite different from the low resolution spectrum published by Hu
et
al. (1993). The derived temperature in the present work of 4250$\pm$200K correspond
more to and early K-type stars and not to an F5 as classified by these authors.

The spectral characteristcs of IRAS~16559-2957 do not correspond with the
properties of O-rich PAGB's. On the other hand many of them compare with
properties of giant stars at the AGB as it has been documented in $\S$\ref{sec:cno}
and $\S$\ref{sec:heavy} and $\S$\ref{sec:discuss}.

 Our results$^{12}$C/$^{13}$C=15 and [C/Fe]= $-0.68$ are consistent with first
dredge up predictions (Iben 1995).  We also note the presence
of Rb, only observed in AGB stars, very few field giant stars and in old open
clusters.

The small photospheric radial velocity of $-2.3 \pm 0.5$ km s$^{-1}$ (V$_{\rm
LSR}$
= +7.1 km s$^{-1}$) is typical of the
thin disk population. It is supported by near solar metallicity measured for this
object. The OH maser V$_{\rm LSR}$  of 57--70 km s$^{-1}$ (te Lintel Hekkert et al.
1991)
correspond to the material being ejected hence red shifted to the observer. Our high
resolution
spectrum, and the low resolution spectrum of Hu et al. (1993) do not exhibit any
emission, hence circumstellar material causing IR flux and maser may be much farther
from the photosphere and cooler. OH masers in AGB stars are common (e.g. Reid et al.
1977; Bowers et al. 1989) and ocurre at distances$> 10^3$ AU from the central star
(Reid \& Moran 1981). No OH maser blue shifted component has been observed. If
this scenario is correct, given the low observed stellar velocity, either the
outflow is not spherically symetric or the blue component should have a velocity of $
\sim -60$ km s$^{-1}$, which would indicate a very large outflow velocity, but this
has been seen in some post-AGB shells, e.g. Zijlstra et al. (2001).

From the atmospheric detailed abundance analysis, we conclude that
IRAS~16559-2957
is experiencing their first crossing towards the red giant branch and has probably 
undergone the first dredge-up. No traces of third dredge up are observed. 
IRAS~16559-2957 is unlikely to be a post-AGB star.

\section*{Acknowledgments}

We are indebted to Prof. D.L. Lambert for permitting the use of the high resolution
spectrum of IRAS~16559-2957 used in this work and to Prof. Sunetra Giridhar for her
comments and suggestions on the abundance analysis. The pertinent comments and 
inputs of an anonymous referee are fully acknowledged. This work was supported
by DGAPA-UNAM project IN104612.

\end{document}